# Effect of interface quality on spin Hall magnetoresistance in Pt/MgFe$_2$O$_4$ bilayers

Masafumi Sugino[1], Kohei Ueda[1,2,3*], Takanori Kida[4], Masayuki Hagiwara[4], and Jobu Matsuno[1,2,3]

[1]*Department of Physics, Graduate School of Science, Osaka University, Osaka 560-0043, Japan*
[2]*Center for Spintronics Research Network, Graduate School of Engineering Science, Osaka University, Osaka 560-8531, Japan*
[3]*Spintronics Research Network Division, Institute for Open and Transdisciplinary Research Initiatives, Osaka University, Yamadaoka 2-1, Suita, Osaka, 565-0871, Japan*
[4]*Center for Advanced High Magnetic Field Science, Graduate School of Science, Osaka University, Osaka 560-0043, Japan*

**Abstract**

We report on spin Hall magnetoresistance (SMR) in bilayers composed of Pt and magnetic insulator MgFe$_2$O$_4$ (MFO) with spinel structure. The Pt thickness dependence of the SMR reveals that annealing of the MFO surface before depositing the Pt layer is crucial for a large SMR with better interface quality. We also found that oxygen pressure during the MFO growth hardly affects the SMR while it influences on magnetic property of the MFO film. Our findings provide important clues to further understanding the spin transport at interfaces containing magnetic insulators, facilitating development of low power consumption devices.

*e-mail: kueda@phys.sci.osaka-u.ac.jp

In the field of spintronics, generation and transmission of spin current have attracted much attention as an avenue for magnetic memory device application[1]. Of particular interest is spin Hall effect (SHE)[2] of non-magnetic metals (NMs) with strong spin-orbit coupling, such as Pt[3–6], Ta[7–9], and W[10,11]. The SHE generates the spin current that can manipulate magnetization of a neighboring magnetic layer in NM/ferromagnet bilayers. Recently, the SHE is found to give rise to the spin-Hall magnetoresistance (SMR), which is one of the intriguing spin-dependent transport phenomena in the bilayers. So far, the SMR has been mainly studied on magnetic insulator yttrium ion garnet ($Y_3Fe_5O_{12}$, YIG) since the magnetic YIG layer does not exhibit electric current shunting that complicates interpretation of spin transport. Through a number of efforts made in Pt/YIG bilayers[10-19], the SMR has been established as one of the powerful ways to estimate spin transport properties.

Spinel ferrites form a class of magnetic insulators with the formula $A^{2+}Fe_2O_4$[20,21], where $A$ is alkaline-earth elements (Mg, Ca) or transition metal elements (Co, Ni). In contrast to the garnets, a wide variety of magnetism is realized by choice of the $A$ element in the spinel ferrites; this motivates us to explore spin transport using the ferrites. Actually, a sizable SMR has been reported in Pt/spinel ferrite bilayers[22–30]. We choose $MgFe_2O_4$ (MFO) in which $Fe^{3+}$ is the only magnetic ion. Crystal structure of MFO is a mixture of normal and inverse spinels; the ratio of the inverse spinel is referred to as an inversion parameter. One of the merits of MFO thin films is that the magnetization is a simple function of the inversion parameter which can be modified by growth conditions such as oxygen pressure. We also focus on quality of the interface between MFO and Pt. In most of the previous reports[10–19,22,23,26,28–30], Pt is ex situ deposited on the magnetic insulators and hence its interface quality is not well defined. According to the presence or absence of annealing of the MFO surface before depositing Pt, we can control the interface quality. Both the magnetization and the interface quality can affect the SMR signal through spin mixing conductance at the interface.

In this letter, we report on the SMR in Pt/$MgFe_2O_4$ bilayers by investigating the effect of the interface quality and the magnetization. We found that the improvement of the interface quality through the annealing process plays a significant role in the enhancement of the SMR with large spin mixing conductance. In contrast, the magnetization controlled by the oxygen pressure does not have important effect on the SMR.

The epitaxial MFO films were grown on MgO(001) substrates by pulsed laser deposition from ceramic MFO target using a KrF excimer laser ($\lambda$ = 248 nm) at 5 Hz. The substrate temperature was fixed at 850°C with the oxygen pressure of 50 and 100 mTorr. In the bulk form, lattice constants of the MgO substrate and the MFO are 0.4213[25] and 0.8394 nm[31], respectively. The lattice mismatch between MFO and the doubled unit cell of MgO is −0.38%; we can expect that coherent epitaxial growth of MFO thin films on the MgO(001) substrate is easily realized. The crystal structure and the thickness of the MFO/MgO films were confirmed by x-ray diffraction. Figures 1(a) shows typical x-ray $2\theta$–$\theta$ diffraction patterns of the film, indicating peaks of MFO(004) film around MgO(002) substrate, while we confirmed that there are no impurity peaks in the wider $2\theta$ range (not shown). The MFO peaks give the out-of-plane lattice constants ($c_{film}$) of 0.8366 nm for 50 mTorr and 0.8381 nm for 100 mTorr, which are consistent with the values in previous reports[32,33]. Oscillations are observed around the MFO(004) reflections, providing total thickness of MFO of 56.4 nm for 50 mTorr and 50.7 nm for 100 mTorr. Figures 1(b) and 1(c) show the corresponding reciprocal space mappings around the MFO(113) reflection, which represent the horizontal axis and vertical axis corresponding to the MFO(110) [or MgO(110)] and MFO(001) [or MgO(001)], respectively. These results clearly indicate that the in-plane lattice constant of the film $a_{film}$ is locked to 0.8426 nm, twice of the MgO lattice constant. Coherently strained films are thus realized by in-plane tensile strain from the substrate. Given that a hypothetical bulk lattice constant $a_{bulk}$ is obtained by the formula $a_{bulk} = \frac{(1-\nu)c_{film} + 2\nu a_{film}}{(1+\nu)}$, where $\nu$ is the Poisson ratio, we deduce $a_{bulk}$ = 0.8394 nm for 50 mTorr and $a_{bulk}$ = 0.8402 nm for 100 mTorr under the assumption of $\nu$ = 0.3. Since it is known that the bulk lattice constant linearly decreases with the inversion parameter[32], we roughly estimate the inversion parameters for the 50 mTorr and 100 mTorr thin film to be 77% and 70%, respectively; the inversion parameter for the 50 mTorr film is larger than that for the 100 mTorr film. Hence, all the diffraction data illustrate that the high-quality MFO films are epitaxially grown on the MgO(001) substrates while the inversion parameter is controlled by the oxygen pressure.

We then measure the saturation magnetization ($M_s$) of the MFO films by superconducting quantum interference device magnetometer. The magnetization curves in Fig. 1(d) were measured at 300 K while sweeping an in-plane magnetic field ranging from +5 to −5 T, confirming room temperature ferromagnetism. The $M_s$ values were determined to be 1.08 $\mu_B$/f.u. (1.36 × 10$^5$ A/m) for 50 mTorr and 1.85 $\mu_B$/f.u. (2.26 ×10$^5$ A/m) for 100 mTorr, similar to the reported values[33,34]; by applying the out-of-plane magnetic field (not shown), we confirmed that both films exhibit the in-plane easy axis of magnetization, providing the effective perpendicular magnetic anisotropy $K_u^{eff}$ = −4.20×10$^4$ J/m$^3$ for 50 mTorr and $K_u^{eff}$ = −1.83×10$^5$ J/m$^3$ for 100 mTorr. The smaller $M_s$ for the 50-mTorr film is explained by the above-mentioned larger inversion parameter since both the $M_s$ and the lattice constant are reduced as the inversion parameter increases[32,34]. The normalized magnetization curves in the inset indicate that the saturation field is smaller for 50-mTorr film. While we speculate that the different saturation field is strongly connected with the inversion parameter as well through the magnetic anisotropy, it has been unclear how the inversion parameter correlates with the magnetic anisotropy in the spinel ferrites; further experimental study is required to clarify that. Here we focus on the SMR measurements, which require that the magnetization is saturated at room temperature. As shown in the inset of the Fig. 1(d), both films can be actually saturated at the maximum magnetic field of 1.35 T in our SMR setup. We accordingly prepared two types of high-quality MFO epitaxial thin films with different magnetizations.

In order to fabricate Pt Hall bar devices on top of the MFO, 1.5-nm-thick Pt layer via shadow mask was deposited by radio frequency magnetron sputtering with base pressure of ~2 × 10$^{-6}$ Pa and Ar working pressure of 0.4 Pa; dimensions of the device are 250 μm in width ($w$) and 625 μm in length ($l$). Note that all the examined samples were prepared by ex-situ Pt deposition. Figure 2(a)–(c) shows a schematic illustration of the Pt Hall bar device with the coordinate system for the angular dependence of MR measurement. The $zy$, $zx$, and $xy$ scans represent directions of external magnetic field ($B$) corresponding to $zy$, $zx$, and $xy$ plane rotations, respectively; the rotation angle is $\theta$ for $zx$ and $zy$ scans, and $\phi$ for $xy$ scan. The $J$ is defined as applied charge current flowing in the $x$ direction. The $R_\mathrm{L}$ and $R_\mathrm{T}$ are defined as resistance for longitudinal and transverse directions, respectively. The measurements were performed at room temperature with $J = 100$ μA for the electrical conductivity ($\sigma$) and 1 mA for the MR.

We describe two types of angle-dependent MR measurements[12,24], namely, longitudinal and transverse MR. The longitudinal MR focuses on the $R_\mathrm{L}$ by rotating sample with a fixed $B = 1.35$ T in three orthogonal planes; the applied $B$ is sufficiently high to saturate magnetization ($M$) in all the planes. The $zy$ scan corresponds to the SMR that is magnetoresistance due to asymmetry between absorption and reflection of spin current generated from the bulk SHE in NM layer[1,2], resulting in a higher resistance at $M//z$ and lower resistance at $M//y$. The $zx$ scan represents the anisotropic magnetoresistance (AMR) originating from the enhanced scattering of conduction electrons from the localized $d$- orbitals in the bulk ferromagnetic metals[35], resulting in higher resistance at $M//x$ and lower resistance at $M//z$. Since we expect no contribution of the AMR due to insulating MFO, the $xy$ scan stems from the SMR, similar to the $zy$ scan. In the same manner as the longitudinal SMR, the transverse MR is performed by measuring $R_\mathrm{T}$ in $xy$ plane, corresponding to the SMR. Therefore, we express longitudinal SMR/AMR and transverse SMR by the general form[12,22];

$$R_\mathrm{L} = R_0 - \Delta R_\mathrm{L} \sin^2 \theta \ (\varphi), \ (1)$$
$$R_\mathrm{T} = \Delta R_\mathrm{T} \sin(2\varphi), \ (2)$$

where $R_0 \equiv R (M \parallel x)$, $\Delta R_\mathrm{L}$ is the longitudinal resistance difference when the $M$ sufficiently saturates along $z$ axis and $y$ ($x$) axis directions for SMR (AMR) with $B$, $\Delta R_\mathrm{T}$ is the transverse resistance difference when the $M$ sufficiently saturates along $z$ axis and $x$ axis directions with $B$.

Figure 2(d) exemplifies the angle dependence of the longitudinal MR in $zy$, $zx$, and $xy$

planes for the 1.5 nm-thick Pt Hall bar device deposited on the 50mTorr-MFO films. The result exhibits apparent MR curves with the same amplitude of $\Delta R_L > 0$ in $zy$ and $xy$ planes, and invisible MR curve with $\Delta R_L \approx 0$ in $zx$ planes. This explains sufficient SMR and negligible AMR, in agreement with the above description. We note that all of the results here exclude the magnetic proximity effect (MPE), which would cause sizable AMR, in accordance with previous studies[27,29,30]. Accordingly, $R_L$ and $\Delta R_L$ indicates the longitudinal SMR contribution hereafter. The angle-dependent transverse MR is also shown in Fig. 2(e), which exhibits clear curve with amplitude of $\Delta R_T > 0$ due to the SMR. Following Eq. (1) and (2), the $\Delta R_{L(T)}$ and $R_0$ were obtained by the fit on the longitudinal and transverse SMR curves in Figs. 2(d) and 2(e). Normalized longitudinal and transverse SMR are defined as L-SMR = $\Delta R_L/R_0$ and T-SMR = $(\Delta R_T \times w/l)/R_0$; the dimensional factor $w/l$ is considered for the transverse SMR, providing L-SMR = 0.067% and T-SMR = 0.046%. Note that while both the SMR is considered to have physically the same origin as found in prior studies[12,22], the larger L-SMR than the T-SMR would be possibly from insufficient accuracy of the dimensions of the Pt-Hall bar structure via the shadow mask; the sputtered Pt particles can be moved into the gap between the mask and the MFO film. In contrast, the above-mentioned studies[12,22] used the Hall-bar structure with well-defined dimensions fabricated by lithography technique. We expect that the ratio of L-SMR is more accurate compared to the T-SMR since the corresponding dimensions are larger for L-SMR.

We examined the SMR as a function of Pt thickness ($t$), where the $t$ was changed from 1.5 to 8 nm. Here, we prepare three bilayer samples for comparison. The bilayers including the MFO films grown under 50 mTorr and 100 mTorr are labeled as 50 mTorr and 100 mTorr, respectively. The bilayer including the MFO film grown under 50 mTorr with annealing process is labeled as 50 mTorr-anneal, where the annealing process was carried out in the base pressure of the sputtering chamber (~$10^{-7}$ Torr) for 1 hour at 400°C before the Pt deposition. The SMR data is displayed in Fig. 3(a) – (c), all of which exhibits the decreased amplitude with increasing $t$ and the maximum amplitude at 1.5 nm. Among three samples, the 50 mTorr-anneal has the largest amplitude ~ 0.09% for L-SMR and ~ 0.07% for T-SMR, suggesting the significant spin injection across the Pt/MFO interface. We estimated the spin mixing conductance $G_r^{\uparrow\downarrow}$ and the spin diffusion length $\lambda$ of Pt layer as described in Refs. 11,12,22,26.

$$\text{L(T)-SMR} = \theta_{SH}^2 \frac{\lambda}{t} \frac{2\lambda G_r^{\uparrow\downarrow} \tanh^2\left(\frac{t}{2\lambda}\right)}{\sigma + 2\lambda G_r^{\uparrow\downarrow} \coth\left(\frac{t}{\lambda}\right)} \quad (3).$$

Assuming the spin Hall angle $\theta_{SH} = 0.10$[10,26] of Pt, we fit these data in Fig. 3 to Eq. (3). The $t$-dependent $\sigma$ was obtained from our experiment with a theoretical model[12,13] since the $\sigma$ of Pt on MFO film is not constant in thinner $t$ range. We confirm that the Pt quality is almost the same in all the Pt/MFO bilayers, indicating that the extracted interface resistivity and bulk mean free path are comparable to the previous studies[36–38]. The SMR data of all the samples is nicely fitted, yielding the values of $G_r^{\uparrow\downarrow}$ and $\lambda$. While these values are comparable to many reported values in the literature of Pt/magnetic insulator bilayers[10–19,26–29], we do not further discuss the $\lambda$ since our focus is the $G_r^{\uparrow\downarrow}$ that greatly influences the SMR; the $\lambda$ ranges from 0.8 to 1.3 nm, consistent with previous reports[3,4,12,19,29]. Here, all the values of $G_r^{\uparrow\downarrow}$ are summarized in Table 1.

The $G_r^{\uparrow\downarrow}$ for the 50 mTorr-anneal is larger by 40% or more than those for the 50 mTorr and 100 mTorr, which are roughly comparable to each other; the similar trend found in L-SMR and T-SMR indeed indicates the validity of our experimental evidences. This enhancement of the SMR demonstrates that the interface of the 50 mTorr-anneal is likely to be improved due to the annealing process since all of the samples involve exposure of the MFO surface to air before Pt deposition. A similar annealing process is reported to be effective in cleaning the Pt/YIG interface without any change of the film crystallinity while the process can remove the surface adsorbate[17]. We therefore expect that the anneal temperature of 400℃ in our case is effective in removing the water molecule while the temperature is too low to affect the crystal structure and/or crystallinity of our MFO films since the stabilization of the crystal structure during the deposition requires the substrate temperature as high as 850℃. In contrast, the similar SMR magnitude between the 50 mTorr and 100 mTorr indicates that the magnetization does not influence on the SMR considering that the oxygen pressure increases the magnetization of the MFO film discussed in Fig. 1(d). It is known that the enhanced SMR in Pt/spinel ferrite $NiFe_2O_4$ (NFO) bilayers with increasing the magnetization, which is obtained by increasing the oxygen pressure during the growth of the NFO films[30]; the result is attributed to be a weak magnetization of the Pt stemming from the MPE. Taking into account that the MPE is experimentally excluded in our case of Pt/MFO as discussed above, our results suggest that the SMR is independent of the magnetization in the absence of the MPE. Thus, our result clearly demonstrates how the spin mixing conductance $G_r^{\uparrow\downarrow}$ correlates with the two significant factors of the interface between NM and magnetic insulators, that is, the interface quality and the magnetization, providing fruitful information to clarify the spin transport based on magnetic insulators. Moreover, a non-magnetic spinel oxide replaced from Pt enables us to grow all oxide epitaxial structure, highlighting that the high-quality interface can be a good platform for pursuing intriguing spin-current physics; one of the suitable candidates is $LiIr_2O_4$[39] that would be efficient spin current source since recent reports have shown large spin current generation arising from $5d$ iridium oxides[40-45] with strong spin-orbit coupling. In addition to the SMR measurement, other approaches such as the spin pumping measurement would provide further understanding of the oxide interface based on the spinel ferrite.

In conclusion, we have studied that the effect of interface quality and magnetization on spin transport, by investigating the longitudinal and transverse SMR of Pt films deposited on epitaxial magnetic insulating MFO films. While the effect of the oxygen pressure is critical in magnetization, it is not significant in the SMR with the similar $G_r^{\uparrow\downarrow}$. As found in the largest $G_r^{\uparrow\downarrow}$ in the 50 mTorr-anneal, the interface quality has the more important effect on spin transport at interface. These results may have profound implications for efficient charge to spin current generation, enabling the design of future spintronic devices based on magnetic insulators.

**Acknowledgement**

We would like to thank T. Arakawa for technical support. This work was carried out at the Center for Advanced High Magnetic Field Science in Osaka University under the Visiting Researcher's Program of the Institute for Solid State Physics, the University of Tokyo. This work was also supported by the Japan Society for the Promotion of Science (JSPS) KAKENHI (Grant Nos. JP19K15434, JP19H05823, and JP22H04478), JPMJCR1901 (JST-CREST), and Nippon Sheet Glass Foundation for Materials Science and Engineering. We acknowledge stimulating discussion at the meeting of the Cooperative Research Project of the Research Institute of Electrical Communication, Tohoku University.

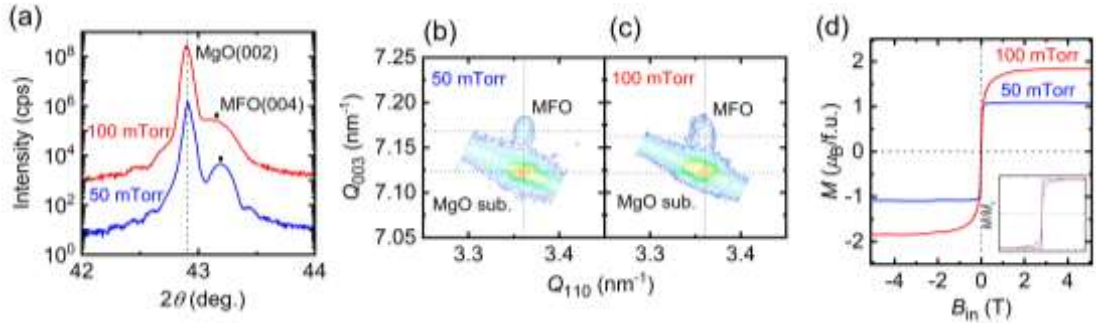

Fig. 1. (a) $2\theta$-$\theta$ scan of MgFe$_2$O$_4$(MFO) grown on MgO substrate with oxygen pressure of 50 mTorr (blue) and 100 mTorr (red) at substrate temperature $T_{sub}$ = 850 °C. Black dot line and black circles indicate the MgO(002) and the MFO(004) orientations, respectively. Reciprocal space mapping around the MFO(113) film peak with oxygen pressure of (b) 50 mTorr and (c) 100 mTorr. (d) Magnetization curves measured at 300 K for the film along the in-plane direction with oxygen pressure of 50 mTorr (blue) and 100 mTorr (red).

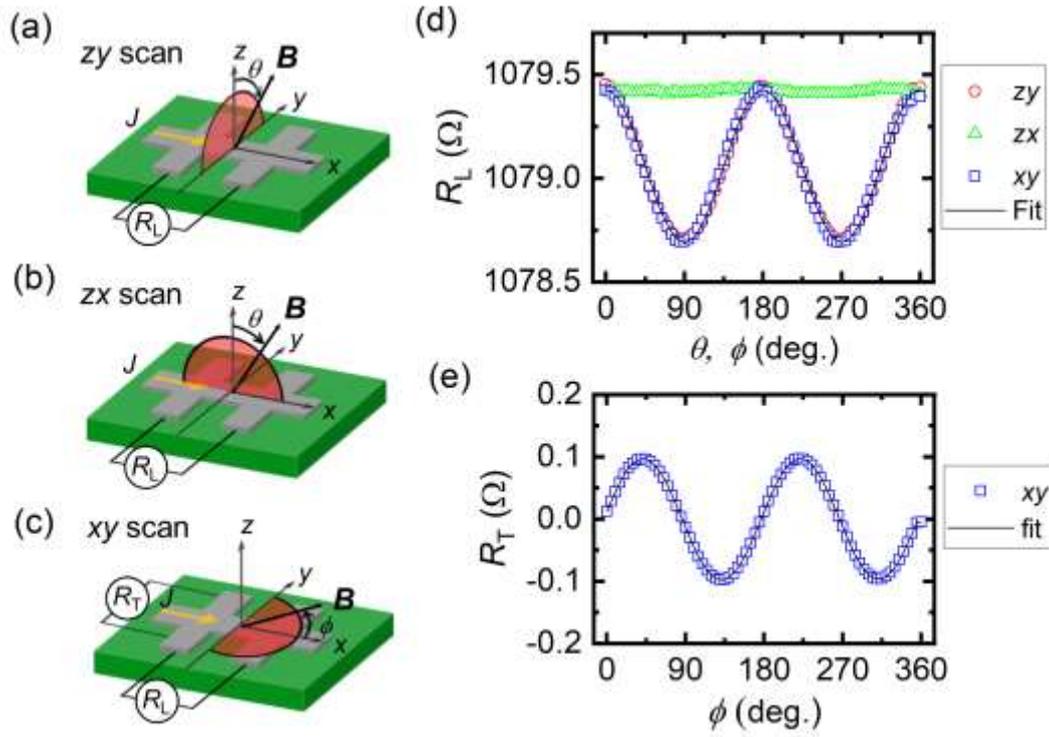

Fig. 2. (a–c) Definition of the coordinate system and schematic illustration of a Pt Hall bar device. The *zy*, *zx*, and *xy* scans represent directions of external magnetic field corresponding to *zy*, *zx*, and *xy* plane rotations, respectively. The $J$ is defined as applied charge current, where it flows in the *x* direction. During the application of the $J$, resistance of the longitudinal and transvers directions ($R_L$, $R_T$) are measured. (d) Angle dependence of the longitudinal magnetoresistance obtained from *zy*, *zx*, and *xy* scans in a Pt(1.5 nm)/MFO/MgO substrate. (e) The corresponding result of the transverse SMR obtained from the *xy* scan.

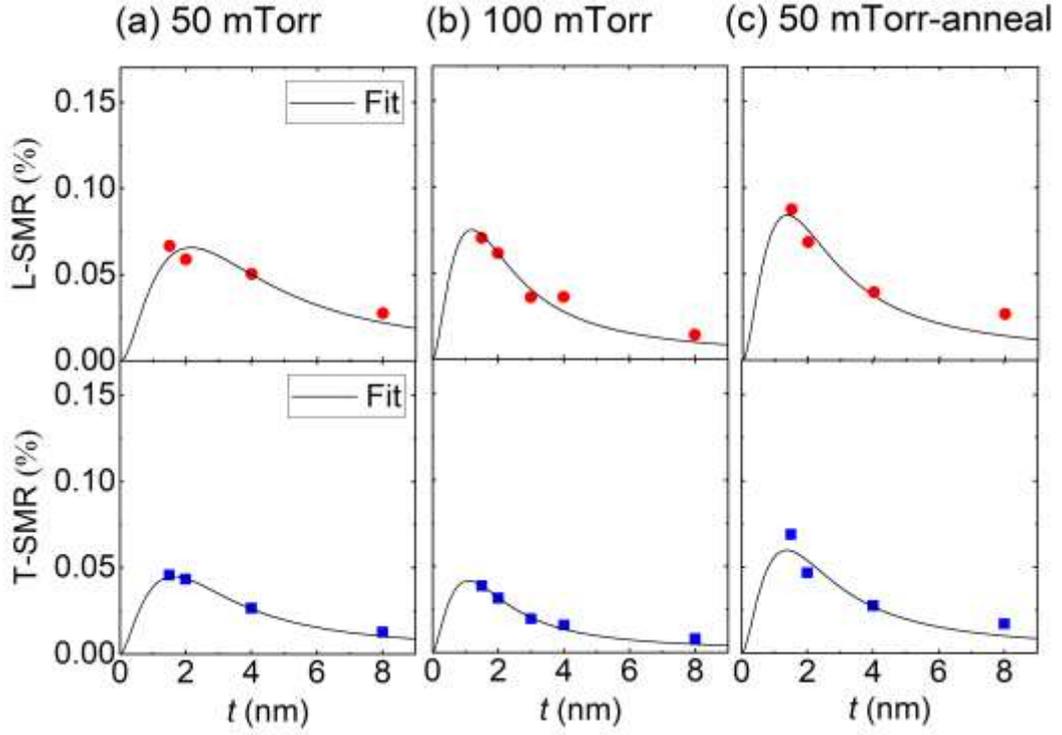

Fig. 3. (a) The longitudinal and transverse SMR ratios as a function of Pt thickness with various conditions: oxygen pressure of (a) 50 mTorr and (b) 100 mTorr without annealing treatment, and (c) that of 50 mTorr with annealing process. Black solid curves denote fitting result based on the SMR model in Eq. (3).

|  | 50 mTorr | 100 mTorr | 50 mTorr-anneal |
| --- | --- | --- | --- |
| L-SMR | 3.5 ± 0.3 | 4.0 ± 0.1 | 5.3 ± 0.1 |
| T-SMR | 2.5 ± 0.3 | 2.7 ± 0.2 | 3.8 ± 0.2 |

Table 1. Summary of the extracted spin mixing conductance, $G_r^{\uparrow\downarrow}$ ($10^{14}$ $\Omega^{-1}$m$^{-2}$), for L-SMR and T-SMR in all the samples.
12